# Spin dynamics slowdown near the antiferromagnetic critical point in atomically thin FePS$_3$

*Xiao-Xiao Zhang[1,2]\*, Shengwei Jiang[3,4], Jinhwan Lee[5], Changgu Lee[5], Kin Fai Mak[2,3,6] and Jie Shan[2,3,6]*

[1]Department of Physics, University of Florida, Gainesville, Florida, USA

[2]Kavli Institute at Cornell for Nanoscale Science, Ithaca, New York, USA

[3]Laboratory of Atomic and Solid State Physics, Cornell University, Ithaca, New York, USA

[4]School of Physics and Astronomy, Shanghai Jiao Tong University, Shanghai, 200240, China

[5]Mechanical Engineering, Sungkyunkwan University, 2066 Seoburo Jangan-gu, Suwon, Gyeonggi-do Korea 16419

[6]School of Applied and Engineering Physics, Cornell University, Ithaca, New York, USA

**ABSTRACT:** Two-dimensional (2D) magnetic materials have attracted much recent interest with unique properties emerging at the few-layer limit. Beyond the reported impacts on the static magnetic properties, the effects of reducing the dimensionality on the magnetization dynamics are also of fundamental interest and importance for 2D device development. In this report, we investigate the spin dynamics in atomically-thin antiferromagnetic $FePS_3$ of varying layer numbers using ultrafast pump-probe spectroscopy. Following the absorption of an optical pump pulse, the time evolution of the antiferromagnetic order parameter is probed by magnetic linear birefringence. We observe a strong divergence in the demagnetization time near the Néel temperature. The divergence can be characterized by a power-law dependence on the reduced temperature, with an exponent decreasing with sample thickness. We compare our results to expectations from critical slowing down and a two-temperature model involving spins and phonons, and discuss the possible relevance of spin-substrate phonon interactions.

The recently emerged two-dimensional (2D) van der Waals magnetic materials offer an attractive platform to explore both fundamental problems in magnetism in the 2D limit and new possibilities in designing tunable spintronics devices[1-13]. Much progress has been made in recent years, including the demonstration of electrical writing and readout of magnetic states [5-6, 8-9, 11-12, 14-16], control of the magnetic states by interlayer stacking [17-18], direct imaging and control of critical spin fluctuations [19], and control of sub-THz spin waves [20-21]. While most efforts have focused on the equilibrium magnetic properties, the ultrafast spin dynamics in these materials have remained largely unexplored. A question of fundamental interest is how the magnetic states of 2D magnetic materials respond to an ultrafast external perturbation. Understanding such non-equilibrium magnetization dynamics will provide important insights into the spin-phonon interactions and spin relaxation mechanisms in these materials.

Ultrafast magneto-optical spectroscopy is a powerful method to address this question. It can directly access the spin dynamics on a wide range of timescales[22-23] and probe the interactions between the different degrees of freedom (e.g., charge, spin, and lattice) in a material [24-25]. It is particularly useful for investigating spin dynamics in exfoliated 2D magnetic crystals of mesoscopic dimensions, where conventional non-local techniques, such as neutron scattering, cannot be readily applied.

In this work, we investigate the ultrafast spin dynamics in atomically thin $FePS_3$ near its antiferromagnetic (AF) critical point. $FePS_3$ is a van der Waals crystal with highly anisotropic AF exchange interactions: the in-plane exchange is substantially stronger than its out-of-plane counterpart [26-27]. The spins are AF-aligned and fully compensated within each monolayer. Because of the weak interlayer coupling, both bulk and few-layer $FePS_3$ can be viewed approximately as a quasi-2D Ising antiferromagnet [28-32]. Neutron scattering experiments on bulk $FePS_3$ have also shown the formation of in-plane spin stripes below the Néel temperature (~ 120 K) [33-34], as illustrated in Fig. 1a.

We measure $FePS_3$ samples of 3-12 layer thickness, obtained by mechanical exfoliation from bulk crystals. Despite the lack of a net sample magnetization, we identify a pronounced optical linear birefringence that can be used as a probe of the AF order parameter. To examine the ultrafast spin dynamics, we use a femtosecond laser pulse above the sample's bandgap (photon energy ~3 eV), which instantaneously excites the electronic subsystem. We probe the subsequent time evolution of the linear birefringence with a time-delayed probe pulse below the bandgap at ~ 1.6 eV. Near the Néel temperature $T_N \sim 110$ K, we observe a strong divergence in the demagnetization time which follows a power-law dependence on the reduced temperature $t = (T - T_N)/T_N$. The result can be qualitatively understood by considering the critical behaviors near the phase transition. The power-law exponent also decreases with sample thickness, suggesting the potential relevance of spin-substrate phonon interactions.

We first characterize the optical linear birefringence in the magnetic state by optical reflection measurements. Linearly polarized light of ~ 1.6 eV photon energy is focused onto the sample at normal incidence, and the reflected intensity $I$ is measured while varying the incident polarization angle. Figure 1b shows the polarization angle dependence of the reflection contrast at 10 K for a 10-layer sample. The reflected light intensity reaches maximum and minimum ($I_{max}$ and $I_{min}$) along two orthogonal directions denoted as $\theta = 0°$ and $90°$, respectively. The normalized intensity $I/I_{min}$ can be fitted to a sinusoidal function of $\theta$ with a $\pi$-period (solid line). No magnetic circular birefringence can be detected, as expected from the zero-net magnetization.

We characterize this linear birefringence by the percentage reflection contrast between these two principal axes $LB = \frac{I_{max} - I_{min}}{I_{max} + I_{min}}$. The photon energy dependence of $LB$ is shown in Fig. 1c. A change in sign occurs near 1.85 eV, reflecting the presence of an optical resonance in FePS$_3$. The detailed spectral dependence as well as the amplitude of $LB$ is dependent on the substrate interference effects[35]. We choose a probe (1.6 eV) well below the resonance, at which the interference effects are not expected to strongly influence the measured dynamics and temperature dependences. For comparison, the SiO$_2$/Si substrate shows no linear birefringence.

Figure 1d shows the temperature dependence of $LB$ at 1.6 eV. With increasing temperature, the linear birefringence decreases sharply near the Néel temperature $T_N \approx 110$ K. Apart from a weak and constant remnant optical anisotropy background above $T_N$ (dashed line), the overall temperature dependence can be fitted to the magnetic order parameter in a 2D honeycomb lattice

Ising model[36] (dotted line). The absence of hysteresis in thermal cycles is consistent with the reported continuous magnetic phase transition in FePS$_3$ [29, 32]. These results support that the linear birefringence signal *LB* can be regarded as an approximate measure of the AF order parameter.

The observation of strong linear birefringence below $T_N$ can be correlated with the formation of spin stripes in FePS$_3$, which spontaneously breaks the rotational symmetry. The result is also consistent with the reported magneto-Raman studies on AF FePS$_3$ [29, 31-32]. Unlike conventional magnetic birefringence (or magneto-optical Voigt effect) [37-38], in which the optical conductivities are different for excitation parallel and perpendicular to the spin orientation, spins are aligned along the out-of-plane direction in FePS$_3$ and parallel to the light propagation direction. The observed linear birefringence here reflects the anisotropic responses parallel and perpendicular to the direction of the spin stripes, which is likely originated from magnetostriction and/or spin-orbit coupling in the material [38]. The remnant optical anisotropy above $T_N$ reflects the anisotropic crystal structure of few-layer FePS$_3$; although the crystal structure of monolayer FePS$_3$ is rotational symmetric, interlayer stacking [29] lowers the rotational symmetry of the system.

With the linear birefringence established as a measure of the AF order parameter, we now proceed to time-resolved studies of the order parameter. We optically pump the FePS$_3$ sample above its bandgap [39] by ~3 eV photons, creating instantaneous heating by generating hot electrons and phonons. The pump-induced magnetization dynamics is measured by the polarization rotation $\delta\varphi$ of a linearly polarized probe beam with photon energy ~1.6 eV at different time delay (Fig. 2a). Figure 2b shows the time evolution of $\delta\varphi$ for three different incident probe polarization angles $\theta$. The pump-induced rotation $\delta\varphi$ shows opposite signs for $\theta = \pm 45°$. No signal is observed when the probe polarization is aligned with the principal axes ($\theta = 0°$ or $90°$). The inset shows the $\sin 2\theta$ dependence of $\delta\varphi$ at 50 ps time delay. The result does not depend on the pump polarization.

The observation is consistent with pump-induced heating and demagnetization in FePS$_3$. In this picture, the pump light heats up the sample instantaneously and weakens the AF order, which results in a reduction of the linear birefringence according to Fig. 1d and gives rise to a $\sin 2\theta$ dependence of $\delta\varphi$ (see Methods). The rotation angle $|\delta\varphi|$ at $\theta = \pm 45\,°$ provides a direct measure of the pump-induced change in the AF order parameter. Below we will focus on $|\delta\varphi|$ measurements at $\theta = 45°$. The picture of thermal-induced demagnetization is further supported by the temperature dependence (Fig. 2c) of the maximum $\delta\varphi$, which reflects the deduction of AF order after the pulse excitation. The temperature dependence of $\delta\varphi$ peaks at $T_N$, at which the magnetic order is most sensitive to small perturbations. The pump-induced temperature rise at different temperatures varies as the specific heat is not constant[40]. The optical pump instantaneously heats up the sample by $\Delta T$ and changes the optical anisotropy $\frac{d\,LB}{dT}$ (Fig. 1d) by $\frac{d\,LB}{dT}\Delta T \approx 2\delta\varphi$ (see Methods). A maximum transient temperature rise ~ 0.1 K near $T_N$ can be estimated. In this study, we limit our measurements within the low pump fluence (50 μW to 150 μW) regime where $\delta\varphi$ scales linearly with the optical pump power and no significant power dependence is observed in the dynamics (see Supporting Information figure S1).

We now examine the temperature dependence of the dynamics. Figure 3a shows the temporal evolution of $\delta\varphi$ (maximum value normalized to 1 for comparison) at three different temperatures, with Fig. 3b zooming in on the initial dynamics. The initial rise in $\delta\varphi$ corresponds to the demagnetization process induced by optical pumping. The subsequent decay is a slow recovery where the quasi-equilibrium systems of magnons and phonons equilibrate with the substrate. The observed dynamics show a strong temperature dependence. In particular, there is an evident slowdown near $T_N$ *in demagnetization*. We can fit the initial rise by exponential growth and the slow recovery by an exponential decay. The obtained temperature dependence of the characteristic rise time $\tau_r$ (the demagnetization time) and the decay time constant $\tau_d$ is plotted in Fig. 3c and Fig. 3d, respectively. We note that the decay time $\tau_d$ is substantially longer than $\tau_r$ at all temperatures so that we can effectively separate the demagnetization and the slow recovery processes. While the increase in decay time near $T_N$ can be attributed to the slower thermal relaxation, the over two orders of magnitude increase in the rise time $\tau_r$ is more intriguing and directly manifests the criticality at the phase transition point. To characterize the divergence in demagnetization, we fit the temperature dependence by a power-law dependence on the reduced temperature $\tau_r \propto |t|^{-m}$ and obtain an exponent $m \approx 1.7 \pm 0.3$ for $T < T_N$ and $m \approx 1.1 \pm 0.3$ for $T > T_N$ ($T_N \approx 109.5$ K). The different exponent for $T < T_N$ and $T > T_N$ is likely caused by the lack of long-range AF order and the emergence of nanoscale domains immediately above $T_N$. Since our optical probe is most sensitive to domains larger than the spot size (~1 μm), we will focus on the $T < T_N$ regime below.

To gain insights into the demagnetization mechanism, we further examine the sample thickness dependence of the spin dynamics. Figure 4a compares the dependence of $\tau_r$ on the reduced temperature $t$ between the 3-layer and 10-layer FePS$_3$ samples. The Néel temperature $T_N \sim 110$ K is nearly thickness independent, showing that the equilibrium magnetic properties of FePS$_3$ remain largely unmodified with reducing thickness. $\tau_d$ also remains substantially longer than $\tau_r$ at all temperatures for all samples. Power-law temperature dependence of $\tau_r$ is seen in both cases. The rise time $\tau_r$ becomes substantially shorter in the thinner 3-layer sample. The extracted exponent $m$ is $0.6 \pm 0.1$ and $1.7 \pm 0.3$ for the 3-layer and 10-layer samples, respectively. We note here that such differences do not come from the differences in pump absorption, as our measurements are carried out in the low-fluence limit, where the dynamics show no obvious power dependence (see Supporting Information Fig. S1). The sample thickness dependence of $m$ is summarized in Fig.

4b. Despite the large uncertainties in $m$, we observe a clear decrease in $m$ with decreasing sample thickness. As the layer number decreases, the demagnetization slowdown becomes less divergent near the critical point.

The non-equilibrium spin dynamics near a magnetic critical point is a challenging problem and has remained not well-understood [41-42]. We compare our results to known mechanisms of demagnetization slowdown near the critical point. The first is the critical slowing-down of spin dynamics, which reflects the divergent spin-spin correlation times at the critical point of a continuous magnetic phase transition. Such a slowdown under thermal equilibrium has been studied both theoretically and experimentally [19, 43-45]. Power-law dependence of the correlation time on the reduced temperature is expected with a dynamic critical exponent dependent on the universality class. In general, the dynamic critical exponent for a particular universality class increases with decreasing dimensionality [44]. The extracted $m \approx 1.7 \pm 0.3$ for thick FePS$_3$ samples is in reasonable agreement with the expected correlation time exponent ~2.1 for a 2D Ising antiferromagnet obtained from Monte Carlo simulations [46] (FePS$_3$ is quasi-2D due to its layered structure). However, the observed decrease in $m$ with sample thickness in Fig. 4b cannot be explained by the critical slowing down picture: thinner samples are closer to the ideal 2D limit so that a larger $m$ would be expected. The disagreement can be understood since the calculated critical slowing down arises from spontaneous spin fluctuations in thermal equilibrium near the critical point, optical pumping in our experiment brings the system out of equilibrium. Here, spin relaxation through coupling to other degrees of freedom (e.g., phonons) becomes important and is beyond the theory of critical slowing down.

To account for non-equilibrium effects, we compare our results to a phenomenological two-temperature model, which considers the heat exchange between the spin and the phonon subsystems. Immediately after the optical pump pulse, hot electrons are created and quickly reach quasi-equilibrium with the phonon subsystem on picosecond timescales. Because the phonon specific heat capacity is larger than the electronic specific heat capacity in a magnetic insulator like FePS$_3$, most of the absorbed energy resides in the phonon subsystem picoseconds after the pump. Since the demagnetization dynamics near the critical point is substantially slower than the initial picosecond relaxation dynamics, we can regard the effect of optical pumping as creating an instantaneous increase in the phonon temperature. Through spin-phonon scattering, the spin subsystem reaches quasi-equilibrium with the phonon subsystem. The relaxation process can be described by a rate equation, $c_s \, dT_s/dt = -g_{sp}(T_s - T_p)$. Here $c_s$ is the specific heat capacity of the spin subsystem, $g_{sp}$ is the spin-phonon coupling rate, and $T_s$ ($T_p$) is the spin (phonon) temperature. Because we have $\tau_d \gg \tau_r$ in our experiment and the phonon specific heat capacity of FePS$_3$ is much bigger than the spin specific heat capacity near the critical point [24-25, 47], we can assume a constant $T_p$. Under this approximation, the rise time $\tau_r$ is simply the thermal RC time constant $c_s/g_{sp}$. If we further assume a weak temperature dependence of $g_{sp}$ near $T_N$, $\tau_r$ is expected to follow the divergence of $c_s$ near the critical point ($c_s \propto \ln|t|$ and $c_s \propto |t|^{-0.125}$ for 2D and 3D Ising models [48], respectively). As shown in Fig. 4a, the temperature dependence of $\tau_r$ of the 3L sample is close to the expected logarithmic divergence behavior for the 2D Ising model (dotted line). However, the layer dependence of the demagnetization dynamics of Fig. 4a cannot be explained by considering the contribution from $c_s$ alone since $c_s$ is expected to become less divergent near the critical point for thicker samples.

Our observation suggests that as the layer number decreases, the dominant factor for demagnetization dynamics may have changed from the spin fluctuations (critical slowing down) to the spin specific heat (through spin-phonon coupling). One plausible explanation is the enhanced substrate phonon coupling to the spin degree of freedom in atomically FePS$_3$. In particular, surface SiO$_2$ phonon modes are known to increase the electronic and spin scattering rates in graphene [49-51]. The spin-phonon interaction strength within the material can also be modified in atomically thin samples due to weakened dielectric screening from the substrate. A comprehensive theoretical study that takes into account the various layer-dependent spin relaxation channels are required to obtain a quantitative description of the experimental observations.

In summary, we have reported magnetic linear birefringence as a probe of the Néel order in few-layer AF FePS$_3$. Using optical pump-probe spectroscopy, we have characterized the divergence behavior of the demagnetization time near the magnetic phase transition. The dependence of the power-law exponent on the sample thickness illustrates the importance of spin-phonon interactions in the 2D limit. Our results demonstrate the importance of the sample-substrate interface, and provide crucial information for developing a comprehensive understanding of spin dynamics and critical phenomena in 2D magnets.

**Methods**

**1. Sample preparation**

Single-crystalline FePS$_3$ was synthesized by the chemical vapor transport (CVT) method. The mixture of elements (99.5% Fe, 99.99% P, and 99.5% S in a molar ratio of 1:1:3; Sigma-Aldrich) was evacuated and sealed in a quartz ampoule. This ampoule was placed into a tubular furnace with the source placed in the middle of the furnace. The furnace was heated up to 700°C at the rate of 1°C/min and was kept at 700°C for 7 days. The ampoule was then cooled down to room temperature. Atomically thin flakes of FePS$_3$ were mechanically exfoliated from bulk crystals onto SiO$_2$/Si substrates (300 nm thick SiO$_2$) and identified by the color contrast under an optical microscope. The layer number of the flakes was further characterized by atomic force microscopy (AFM). For thin samples (< 5 layer), a protective hexagonal boron nitride (hBN) layer was subsequently transferred on top of FePS$_3$ using the dry transfer method [52-53] in order to minimize laser degradation.

## 2. Linear birefringence

To extract the optical anisotropy of the material, we measured the reflection intensity of linearly polarized light while changing the angle between the polarization axis and the crystal axis. A laser beam with tunable photon energy (Coherent Chameleon compact OPO) was focused by a high numerical aperture (~ 0.6) objective onto the sample with a spot size ~ 1 μm. The samples were loaded into a Montana optical cryostat with variable temperatures. A half-wave plate was used to rotate the incident light polarization. The reflected light was collected by the same objective and detected by a balanced photodiode.

We can relate the measured linear birefringence to the material's anisotropic optical response characterized by an optical conductivity tensor $\begin{bmatrix} \sigma_{max} & 0 \\ 0 & \sigma_{min} \end{bmatrix}$. Here $\sigma_{max}$ and $\sigma_{min}$ are the optical conductivity along the two orthogonal principal axes. Consider light reflection from the sample on an isotropic substrate. In the 2D limit, the sample's contribution to the light reflection (at photon energies below bandgap) can be treated as a perturbation. In the main text, we characterized the linear birefringence by the percentage reflection contrast between the two principal axes $LB = \frac{I_{max} - I_{min}}{I_{max} + I_{min}} \approx 2Re[L(\sigma_{max} - \sigma_{min})]$, where $I_{max}$ and $I_{min}$ are the reflection intensities along the two orthogonal principal axes, $L$ is a constant that depends on the local field factor from the structure, and the linear birefringence of the material is assumed to be small ($\sigma_{max} \approx \sigma_{min}$) as in FePS$_3$.

More complete discussion on the relation between electronic anisotropy and polarization rotation in the 2D limit can be found in the Supplementary Information of Ref.54[54]

**3. Time-resolved linear birefringence spectroscopy**

In the time-resolved measurements, we pumped the material with the second-harmonic output (~ 3 eV) of a Ti:Sapphire laser (78 MHz, 200 fs pulse duration from Coherent Chameleon), and probed the linear polarization rotation at ~1.6 eV. The probe photon energy is below the bandgap of the material and no optical excitations are created. The optical setup is similar to the one used for linear birefringence measurements described above. We used ~ 150 μW for the pump beam (spot size ~ 1.5 μm) and ~ 50 μW for the probe beam (spot size ~ 1 μm). The pump beam was linearly polarized and its intensity was modulated at 100 kHz by a combination of half-wave plate, photoelastic modulator and a linear polarizer that is cross-polarized with respect to the original polarization. No dependence on the pump polarization was observed. A half-wave plate was placed in front of the objective to rotate the relative angle between the linear polarization of the probe beam and the sample crystal axis. In the detection path, the pump beam was filtered out by a color filter and the reflected probe beam was collected to go through a half-wave Fresnel Rhomb and a Wollaston prism, and directed onto a balanced photo-detector. The pump-induced change in polarization rotation of the probe beam, i.e., the signal directly from the balanced photo-detector, was detected by a lock-in amplifier at the modulation frequency of the pump.

The pump beam modifies the material's response to $\begin{bmatrix} \sigma_{max} - \delta/2 & 0 \\ 0 & \sigma_{min} + \delta/2 \end{bmatrix}$. When the incoming probe light polarization forms an angle $\theta$ with respect to the $\sigma_{max}$ axis, the polarization of the reflected beam rotates to $\theta'$ after interaction with the sample. In the limit $\delta \ll \sigma_{max} \approx \sigma_{min}$, we have the rotation angle $\delta\varphi = \theta' - \theta \approx \frac{1}{|1+L\sigma_{min}|^2}Re[L\delta]\sin(2\theta)$, which is proportional to the pump-induced change in linear birefringence, and follows a $\sin 2\theta$ dependence.

FIGURES

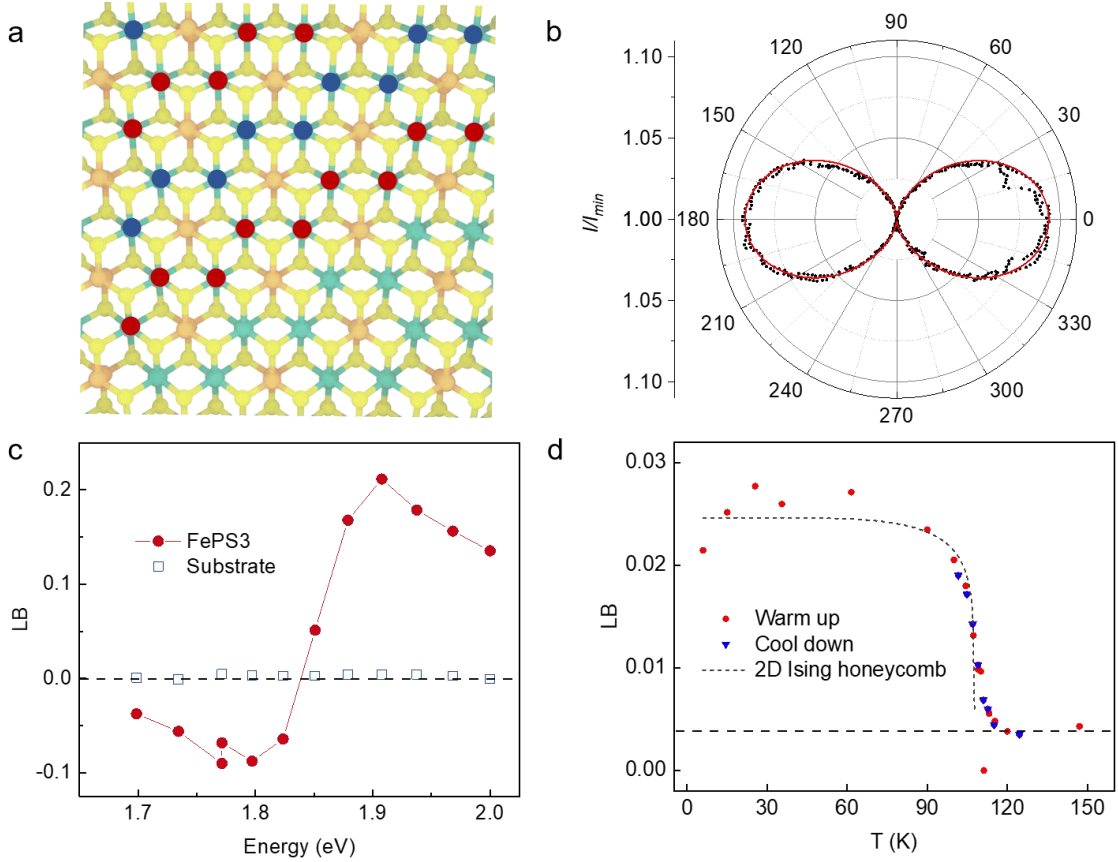

**Figure 1** (a) Atomic structure of FePS$_3$ viewed along the *c* axis. Green, orange and yellow spheres are Fe, P and S atoms, respectively. The formation of spin stripes below the Néel temperature, with up and down spins denoted by blue and red spheres (in the upper left corner), is also shown. (b) Dependence of the reflection contrast $\frac{I}{I_{min}}$ at 1.6 eV on the incident light polarization angle. Zero degree is defined to correspond to the maximum intensity. The measurement temperature is 10 K. The curve is a sinusoidal fit to the data. (c) Linear birefringence (LB) of the sample (red) and the substrate (blue) measured at 4 K as a function of photon energy. (d) Linear birefringence at 1.6 eV as a function of sample temperature from a 10-layer thick sample. Data from warm-up and cool-down cycles are plotted as red dot and blue triangles, respectively. The data is fitted to the magnetic order parameter in a honeycomb-lattice 2D Ising model[36] (dotted line). The horizontal dashed line corresponds to the background lattice anisotropy that is present even above $T_N$.

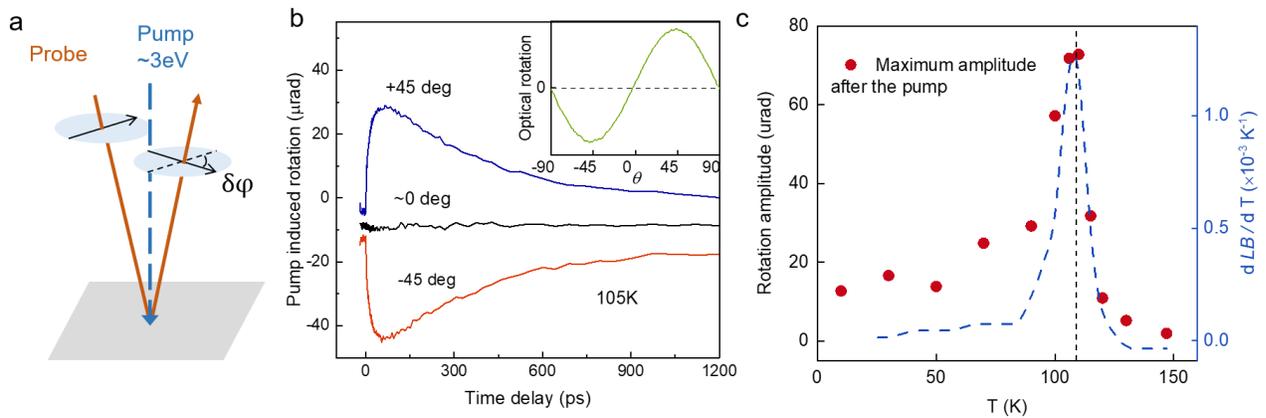

**Figure 2** (a) Schematics of the pump-probe setup with normal incidence geometry. Pump-induced change in the linear birefringence at 1.6 eV is detected by the polarization rotation angle δφ of the time-delayed probe beam. (b) Time-resolved optical rotation of the probe beam for incident polarization angle of the probe beam $\theta$ = +45° (blue), 0° (black) and -45° (red). The angle $\theta$ is measured from the principal axis of maximum/minimum reflectance. The inset shows that δφ at 50 ps time delay follows the sin$2\theta$ dependence. Measurement temperature is 105 K, near the Néel temperature. (c) The maximum probe beam rotation δφ ($\theta$ = 45°) as a function of the sample temperature. For comparison, the blue dashed line is the temperature derivative of the linear birefringence in Fig. 1d. The vertical dashed line denotes the Néel temperature ~110 K. Results are from a 10-layer thick sample.

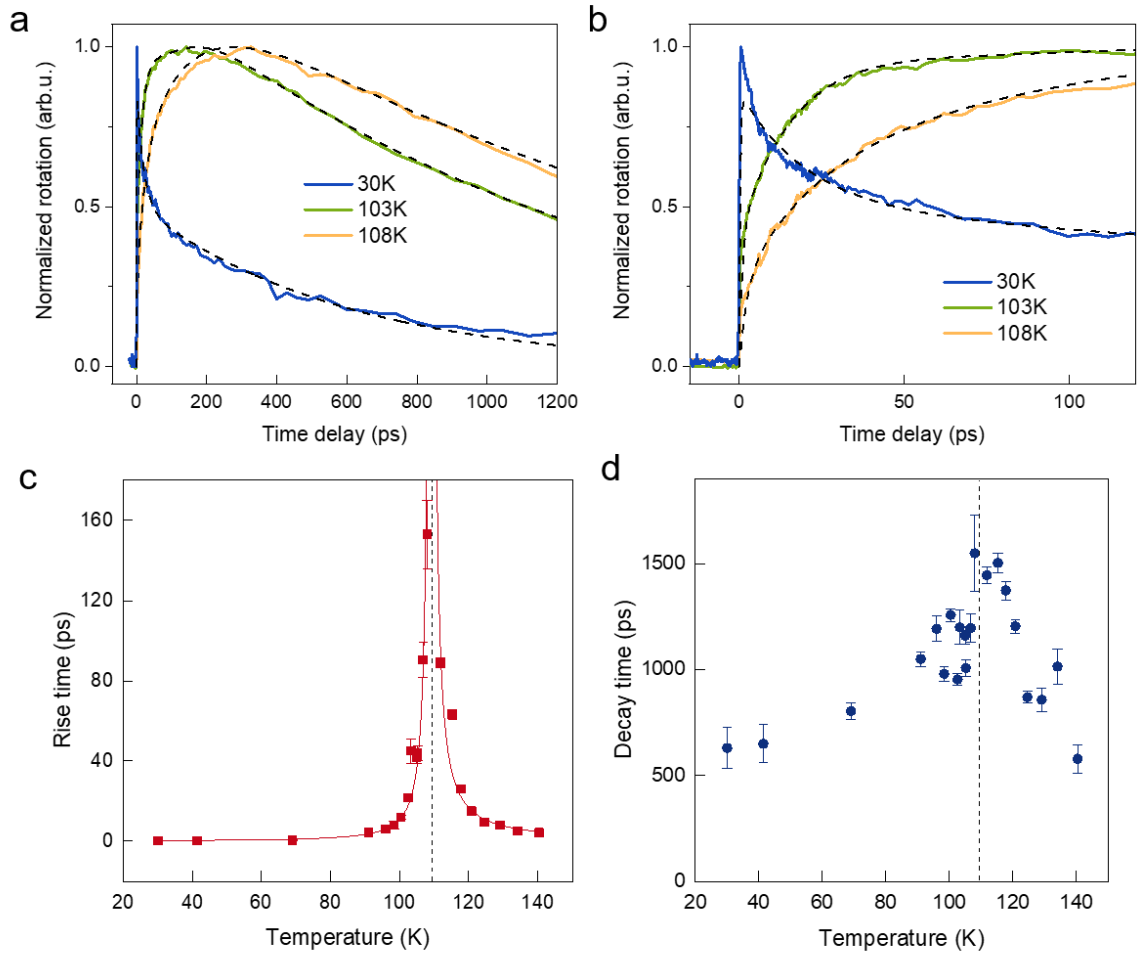

**Figure 3** (a) Time-resolved optical rotation of the probe beam at +45° incident polarization at 30 K, 105 K and 108 K. (b) Zoomed-in time traces of (a) for the first 130 ps. The dashed lines are exponential fits to the data. (c) Extracted rise time as a function of temperature. The solid lines are fits using the power-law dependence on the reduced temperature. The fitted Néel temperature is $T_N$ = 109.5 ± 0.5 K (vertical dashed line). (d) The fitted decay time as a function of temperature, with the vertical dashed line marking the $T_N$ obtained in (c). The decay time is at least an order of magnitude longer than the rise time. Results are from a 10-layer thick sample. The error bars in (c) and (d) are the uncertainties of the exponential fits in (a) and (b).

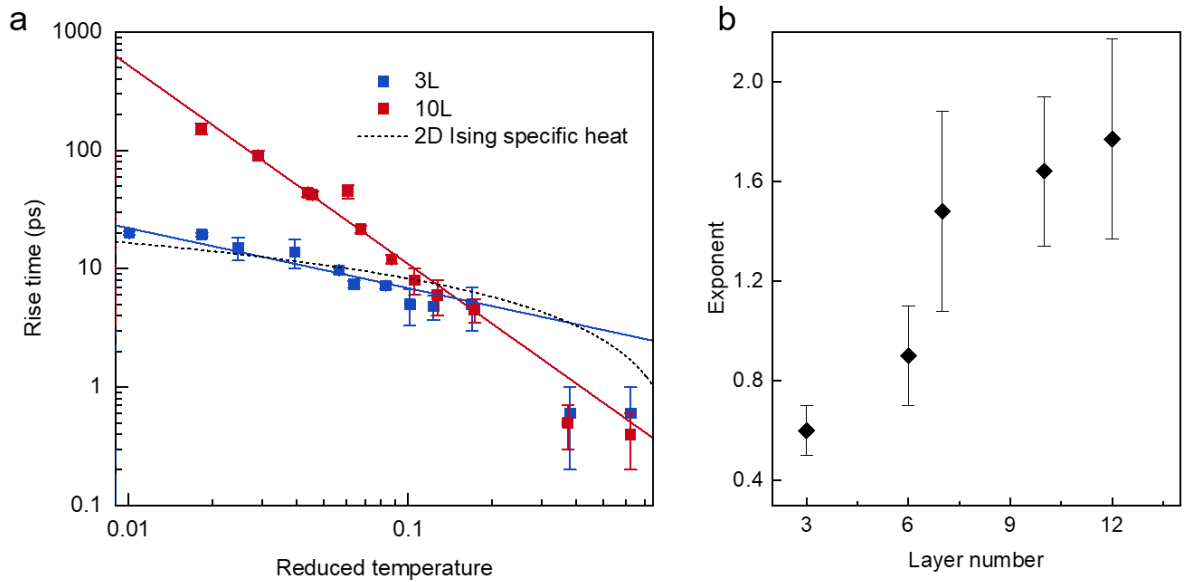

**Figure 4** (a) The extracted rise time of the 3-layer (blue) and 10-layer (red) samples as a function of the reduced temperature (below $T_N$) in a log-log plot. The solid red (blue) line is a power-law fit with exponent 1.7±0.3 (0.6±0.1). The dotted line is the fit of the 3L sample data to the specific heat in the 2D Ising model, which corresponds to a logarithmic dependence. (b) The extracted exponents as a function of sample thickness.

AUTHOR INFORMATION

**Corresponding Author**

*Email: xxzhang@ufl.edu

**Author Contributions**

X.Z., K.F.M and J.S. designed the study. X.Z. developed the time-resolved spectroscopy, performed the measurement and data analysis. S.J. fabricated the samples for measurements. J.L.

and C.L. grew the bulk crystals. All authors discussed the results and commented on the manuscript.